\documentclass[letter,aps,showpacs,superscriptaddress,nofootinbib,tightenlines,twocolumn,prl]{revtex4}

\usepackage{amsfonts}
\usepackage{multirow}
\usepackage{mathrsfs}
\usepackage{graphicx}
\usepackage{amsmath}
\usepackage{amssymb}
\usepackage{bm}
\usepackage{bbm}
\usepackage{color}

\def\OMIT#1{}

\newcommand{\nn}{\nonumber}

\newcommand{\beq}{\begin{equation}}
\newcommand{\eeq}{\end{equation}}
\newcommand{\bqa}{\begin{eqnarray}}
\newcommand{\eqa}{\end{eqnarray}}

\begin{document}
\title{\mbox{}\\[10pt]
Can NRQCD explain the $\gamma\gamma^* \to \eta_c$ transition form factor data?
}

\author{Feng Feng\footnote{F.Feng@outlook.com}}
\affiliation{China University of Mining and Technology, Beijing 100083, China\vspace{0.2cm}}

\author{Yu Jia\footnote{jiay@ihep.ac.cn}}
\affiliation{Institute of High Energy Physics and Theoretical Physics Center for
Science Facilities, Chinese Academy of
Sciences, Beijing 100049, China\vspace{0.2cm}}
\affiliation{Center
for High Energy Physics, Peking University, Beijing 100871,
China\vspace{0.2cm}}

\author{Wen-Long Sang\footnote{wlsang@ihep.ac.cn}}
 \affiliation{School of Physical Science and Technology, Southwest University, Chongqing 400700, China\vspace{0.2cm}}
 \affiliation{State Key Laboratory of Theoretical Physics, Institute of Theoretical Physics, Chinese Academy of Sciences, Beijing 100190, China\vspace{0.2cm}}

\date{\today}
\begin{abstract}
Unlike the bewildering situation in the $\gamma\gamma^*\to \pi$ form factor,
a widespread view is that perturbative QCD can decently account for the recent
\textsc{BaBar} measurement of $\gamma\gamma^*\to \eta_c$
transition form factor.
The next-to-next-to-leading order (NNLO) perturbative
correction to the $\gamma\gamma^*\to \eta_{c,b}$ form factor,
is investigated in the NRQCD factorization framework for the first time.
As a byproduct, we obtain by far the most precise order-$\alpha_s^2$
NRQCD matching coefficient for the $\eta_{c,b}\to \gamma\gamma$ process.
After including the substantial negative order-$\alpha_s^2$
correction, the good agreement between NRQCD prediction and
the measured $\gamma\gamma^*\to \eta_c$ form factor
is completely ruined over a wide range of momentum transfer squared.
This eminent discrepancy casts some doubts on the applicability of NRQCD
approach to hard exclusive reactions involving charmonium.
\end{abstract}



\maketitle

The advent of high-luminosity $e^+e^-$ colliders affords an invaluable
opportunity to study
hard exclusive processes involving light mesons such as $\pi,\eta,\eta^\prime$.
New measurements constantly emerge, some of which greatly
challenge our understanding toward perturbative QCD (pQCD).
In 2009, \textsc{BaBar} collaboration reported their measurement of the $\gamma\gamma^*\to \pi^0$ form factor for
$4\;{\rm GeV^2} <Q^2 < 40\;{\rm GeV^2}$~\cite{Aubert:2009mc}.
Surprisingly, the measured value of $Q^2 |F(Q^2)|$ increases with $Q^2$, exceeding
the asymptote $\sqrt{2}f_\pi=0.185$ GeV predicted in pQCD~\cite{Lepage:1980fj} after $Q^2 > 10\;{\rm GeV^2}$.
This triggered renewed theoretical interest in scrutinizing the leading-twist
pion light-cone distribution amplitude (LCDA), and it looks ironical that to date
our knowledge of pion is still rather inadequate.

Apart from hard exclusive reactions involving light hadrons,
exclusive charmonium production also emerge as a new frontier of pQCD
in recent years~\cite{Brambilla:2010cs}.
This topic was initiated by the double-charmonium production process $e^+e^-\to J/\psi+\eta_c$
first observed by \textsc{Belle} in 2002~\cite{Abe:2002rb}.
Disquieting discrepancy between the data and the leading order (LO)
pQCD predictions~\cite{Braaten:2002fi,Liu:2002wq,Hagiwara:2003cw}
has stimulated a great amount of theoretical activity.

In this work, we are interested in examining
the simplest exclusive charmonium production process:
the $\gamma^*\gamma \to \eta_c$ transition form factor, which is defined through
\beq
   \langle \eta_c (p)\vert J_{\rm EM}^\mu \vert \gamma(k,\varepsilon) \rangle = i e^2 \epsilon^{\mu\nu\rho\sigma} \varepsilon_\nu q_{\rho} k_{\sigma} F(Q^2),
\label{form:factor:def}
\eeq
where $J_{\rm EM}^\mu$ is the electromagnetic current, and the momentum transfer squared $q^2 =(p-k)^2\equiv -Q^2<0$.

In 2010 \textsc{BaBar} collaboration presented very precise measurements for
the $\gamma^*\gamma\to \eta_c$ transition form factor in the range
$2\;{\rm GeV^2}<Q^2<50\;{\rm GeV^2}$~\cite{Lees:2010de}.
The measured $F(Q^2)$ can be well described by a simple monopole fit:
\beq
  |F(Q^2)/ F(0)|= {1\over 1+Q^2/\Lambda},
\label{form:factor:ratio:fit}
\eeq
with the pole parameter $\Lambda=8.5\pm0.6\pm 0.7\; {\rm GeV^2}$.
Interestingly, a variety of phenomenological studies, {\it i.e.},
$k_\perp$ factorization~\cite{Feldmann:1997te,Cao:1997hw},
lattice QCD~\cite{Dudek:2006ut},
$J/\psi$-pole-dominance~\cite{Lees:2010de},
QCD sum rules~\cite{Lucha:2012ar},
light-front quark model~\cite{Geng:2013yfa},
all yield predictions compatible with the data, at least in the small $Q^2$ range.

However, we caution that the satisfactory agreement between the
existing predictions and data should not be interpreted as a triumph of pQCD.
On the contrary,
as we will see later, understanding the $\gamma^*\gamma\eta_c$
form factor actually poses an outstanding challenge to pQCD.

Both light-cone approach~\cite{Lepage:1980fj} and nonrelativisic QCD (NRQCD)
factorization~\cite{Bodwin:1994jh} are well-founded pQCD methods to tackle
hard exclusive reactions involving charmonium.
The bulk of $\eta_c$  events produced in two-photon fusion
concentrates in the  range $Q^2 \sim m^2$,
where light-cone approach is obviously inappropriate.
Nevertheless, since the charm quark propagator
carries typical
virtuality of order $m^2\gg \Lambda^2_{\rm QCD}$
even in the $Q\to 0$ limit, the NRQCD approach, which is based on effective field theory and fully exploits the
nonrelativistic nature of quarkonium,
remains as a valid tool.
As $Q \gg m$, both light-cone and NRQCD approaches can apply, but
lack of precise knowledge of
charmonium LCDA severely hinders the predictive power of the former approach.
In contrast, NRQCD approach is much more predictive, since the only nonperturbative
inputs are a few NRQCD matrix elements that are strongly constrained by the
measured charmonium annihilation decay rates.
The consensus now is that the NRQCD approach generally yields
less ad-hoc, more reliable predictions~\cite{Bodwin:2006dm,Brambilla:2010cs},
in particular amenable to systematically incorporating
perturbative and relativistic corrections.

The aim of this work is to investigate higher-order perturbative corrections
to the $\gamma^*\gamma\to\eta_c$ form factor in the NRQCD framework.
This is motivated by the fact that QCD radiative corrections
in exclusive charmonium  processes are often substantial,
exemplified by the sizable next-to-leading order
(NLO) QCD correction to $e^+e^-\to J/\psi+\eta_c$~\cite{Zhang:2005cha,Gong:2007db,Bodwin:2007ga},
and substantial next-to-next-to-leading order (NNLO) perturbative
corrections to a few $S$-wave quarkonium decay channels, {\it e.g.},
$J/\psi\to e^+e^-$~\cite{Czarnecki:1997vz,Beneke:1997jm},
$\eta_c\to \gamma \gamma$~\cite{Czarnecki:2001zc},
and $B_c\to l\nu$~\cite{Onishchenko:2003ui,Chen:2015csa}.
To the best of our knowledge,
this work constitutes the first comprehensive NNLO analysis for exclusive
quarkonium {\it production} process.

According to the NRQCD factorization~\cite{Bodwin:1994jh},
the form factor at LO in $v$ can be expressed as
\beq
\label{NRQCD:factorization}
    F(Q^2) =  C(Q,m,\mu_R,\mu_\Lambda)\,
    {\langle \eta_c \vert \psi^\dagger \chi(\mu_\Lambda) \vert 0 \rangle \over \sqrt{m}}+{\mathcal O}(v^2).
\eeq
$C(Q,m,\mu_R,\mu_\Lambda)$ represents the NRQCD short-distance coefficient, where $m$, $\mu_R$,  $\mu_\Lambda$ denote the charm quark pole mass, the renormalization scale,
and the NRQCD factorization scale, respectively.
In phenomenological studies, this nonpertubative matrix element is often substituted
by $\sqrt{2N_c} \psi_{\eta_c}(0)$ ($N_c=3$ is the number of color), where $\psi_{\eta_c}(0)$ is the Schr\"{o}dinger
wave function at the origin for $\eta_c$ in quark potential model. In NRQCD,
this matrix element becomes a scale-dependent quantity.

Since the vacuum-to-$\eta_c$ matrix element cancels in the ratio $F(Q^2)/F(0)$,
the NRQCD prediction to (\ref{form:factor:ratio:fit})
is free from any adjustable nonperturbative inputs.
In some sense, this normalized $\eta_c$ form factor bears some similarity with the ratio $\Gamma[J/\psi\to e^+e^-]/\Gamma[\eta_c\to \gamma\gamma]$, which has been carefully studied in \cite{Czarnecki:2001zc,Kiyo:2010jm}.
However, for the latter ratio, the NRQCD matrix elements of $J/\psi$ and $\eta_c$ do not exactly cancel
due to the presence of the spin-spin interaction, the estimate of which requires the
nonperturbative ansatz of modelling the charmonium.
In this respect, confronting the \textsc{BaBar}
data of the $\eta_c$ form factor over a wide range of $Q^2$ would pose a unique and more nontrivial
test against the prediction from NRQCD factorization.

Owing to the fact $Q,m\gg \Lambda_{\rm QCD}$,
the short-distance coefficient can be reliably computed in perturbation theory.
Through $O(\alpha^2_s)$, it can be organized as
\bqa
\label{NRQCD:short-dist:coef:2:loop}
&& C(Q, m,\mu_R,\mu_\Lambda) = C^{(0)}(Q,m) \Bigg\{ 1 + C_F {\alpha_s(\mu_R)\over \pi} f^{(1)}(\tau)
\nn\\
&& + {\alpha_s^2\over \pi^2} \bigg[ {\beta_0\over 4} \ln\! {\mu_R^2
\over Q^2+m^2} C_F f^{(1)}(\tau)
- \pi^2 C_F \left(C_F + \frac{C_A}{2} \right)
\nn\\
&& \times \ln {\mu_{\Lambda}\over m} + f^{(2)}(\tau) \bigg]+{\cal O}(\alpha_s^3)\Bigg\},
\eqa
where $\tau \equiv Q^2/m^2$. $C_F={4\over 3}$ and $C_A=3$ denote the $SU(3)$ color factors.

$C^{(0)}$ signifies the LO NRQCD matching coefficient:
\beq
C^{(0)}(Q,m) = {4 e_c^2\over Q^2+4m^2},
\eeq
where $e_c={2\over 3}$ ($e_b=-{1\over 3}$).
As expected, $C^{(0)}\propto 1/Q^2$ asymptotically.
The LO NRQCD prediction to $F(Q^2)$ coincides with the monopole form in (\ref{form:factor:ratio:fit}),
with the respective pole parameter $\Lambda=4m^2\approx 9\;{\rm GeV}^2$.

The ${\mathcal O}(\alpha_s)$ contribution is encapsuled in $f^{(1)}(\tau)$:
\bqa
&&  f^{(1)}(\tau)  =   {\pi^2(3-\tau)\over 6
(4+\tau)} -{20+9\tau\over 4(2+\tau)} -
{\tau(8+3\tau)\over 4(2+\tau)^2} \ln\!{4+\tau\over 2}
\nn\\
&& +3\sqrt{\tau\over 4+\tau} \tanh^{-1}\!\!\!\!\sqrt{\tau\over 4+\tau}
+{2-\tau\over 4+\tau} \left( \tanh^{-1}\!\!\!\!\sqrt{\tau\over 4+\tau} \right)^2
\nn\\
&&
-{\tau\over 2(4+\tau)}{\rm Li}_2\left(-{2+\tau\over 2}\right),
\eqa
which can be translated from Eq.~(A3) of \cite{Sang:2009jc}.

In Eq.~(\ref{NRQCD:short-dist:coef:2:loop}), $\beta_0 = {11\over 3}C_A - {4\over 3} T_F (n_L+n_H)$ is
the one-loop $\beta$-function coefficient,
$n_H=1$, and $n_L$ labels the number of light quark flavors ($n_L=3$ for $\eta_c$, $4$ for $\eta_b$), and $T_F={1\over 2}$.
The occurrence of the $\beta_0\ln \mu_R$ term in (\ref{NRQCD:short-dist:coef:2:loop})
is dictated by renormalization-group invariance.
The occurrence of the factorization scale $\mu_\Lambda$ reflects that the
NRQCD pseudoscalar density $\psi^\dagger \chi$ has a non-vanishing anomalous
dimension starting at order-$\alpha_s^2$,
as first discovered in \cite{Czarnecki:2001zc}.
The nontrivial task is to decipher the function $f^{(2)}(\tau)$.

We start with computing the on-shell quark amplitude for
$\gamma^*\gamma\to c\bar{c}({}^1S_0^{(1)})$ through order $\alpha_s^2$,
using the covariant trace technique to facilitate the projection of the $c\bar{c}$ pair onto
the intended quantum number.
Prior to performing the loop integration, we neglect the relative momentum between $c$ and $\bar{c}$,
which amounts to directly extracting
the short-distance coefficients at $v^0$ accuracy~\cite{Czarnecki:1997vz,Beneke:1997jm}.

\begin{figure}[hbt]
\begin{center}
\includegraphics[scale=0.5]{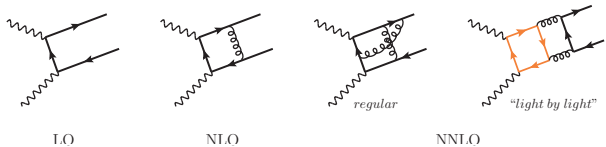}
\caption{Sample Feynman diagrams for $\gamma^*\gamma\to c\bar{c}({}^1S_0^{(1)})$.
\label{feyn:diag}}
\end{center}
\end{figure}

We briefly outline the calculation.
We employ the package \textsf{FeynArts}~\cite{Hahn:2000kx} to generate corresponding Feynman diagrams and
amplitudes through order-$\alpha_s^2$ in Feynman gauge.
There are 108 regular 2-loop diagrams and
12 ``light-by-light" scattering diagrams, two of which are sketched in Fig.~\ref{feyn:diag}.
The latter gauge-invariant subset is UV- and IR-finite.
Dimensional regularization (DR) is utilized to regularize both UV and IR divergences.
We then use \textsf{FeynCalc/FormLink}~\cite{Mertig:1990an,Feng:2012tk} to carry out the trace over Dirac and color matrices.
We follow \cite{Kniehl:2006qw} on the prescription of $\gamma_5$ in DR.
The packages \textsf{Apart}~\cite{Feng:2012iq} and \textsf{FIRE}~\cite{Smirnov:2014hma} are utilized to conduct partial fraction
and integration-by-parts reduction. Finally, we end up with the 174 Master Integrals (MI).
For a handful of simpler MIs, we employ the $\alpha$ parameters~\cite{Smirnov:2012gma} as well as
the Mellin-Barnes tools \textsf{AMBRE}~\cite{Gluza:2007rt}/\textsf{MB}~\cite{Czakon:2005rk} to
infer the (semi)-analytic expressions, while for the remaining MIs,
we combine \textsf{FIESTA}/\textsf{CubPack}~\cite{Smirnov:2013eza,CubPack} to carry out
sector decomposition and subsequent numerical integrations with quadruple precision.

In implementing the quark wave function and mass renormalization,
we take the order-$\alpha_s^2$ expressions of $Z_2$ and $Z_m$ from \cite{Broadhurst:1991fy, Melnikov:2000zc}.
The strong coupling constant is renormalized to one-loop order under $\overline{\rm MS}$ scheme.
All the UV divergences are removed after the renormalization procedure.
At this stage, the NNLO amplitude still
contains a single IR pole with the very coefficient as anticipated from (\ref{NRQCD:short-dist:coef:2:loop}).
We factorize this IR pole into $\eta_c$-to-vacuum NRQCD matrix element according to $\overline{\rm MS}$ prescription,
with $\ln \mu_\Lambda$ manifested as the remnant.
Ultimately, we are able to identify
each individual term as specified in (\ref{NRQCD:short-dist:coef:2:loop}),
with high numerical precision. This is the first calculation that endorses
the validity of NRQCD factorization at the NNLO level for exclusive quarkonium {\it production} process.

\begin{figure}[hbt]
 \includegraphics[width=0.45\textwidth]{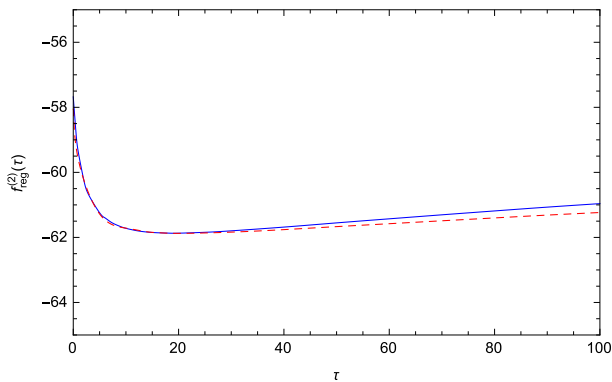}
    \caption{Profile of $f^{(2)}_{\rm reg}(\tau)$. The solid curve for $\eta_c$ ($n_L=3$),
   and the dashed one for $\eta_b$ ($n_L=4$).
   \label{f2}}
\end{figure}

It is convenient to divide the NNLO contribution $f^{(2)}(\tau)$
in (\ref{NRQCD:short-dist:coef:2:loop}) into the regular piece and the ``light-by-light" piece,
$
f^{(2)}(\tau)= f^{(2)}_{\rm reg}(\tau)+f^{(2)}_{\rm lbl}(\tau).
$
The former is real-valued, and the latter is complex-valued,
whose significance and computational cost explode with increasing $\tau$.
$f^{(2)}_{\rm reg}(\tau)$, as plotted in  Fig.~\ref{f2}, turns to be deeply negative,
signaling a substantial negative NNLO correction to the form factor.
Despite being a rather flat function of $\tau$, the asymptote of $f^{(2)}_{\rm reg}(\tau)$ appears to be
numerically consistent with the 2-loop leading collinear logarithm $\ln^2 \tau$ predicted in \cite{Jia:2008ep}.
Some benchmark values of $f^{(2)}_{\rm reg}(\tau)$ and $f^{(2)}_{\rm lbl}(\tau)$
are listed in Table~\ref{Table:1}.
In the experimentally accessible range of $Q^2$,
the ``light-by-light" contribution for $\eta_c$
appears insignificant relative to the regular one.
Nevertheless, the ``light-by-light" piece turns out to be non-negligible
even at relatively low $\tau$ for $\eta_b$.

\def\D{\displaystyle}
\begin{table}[htb]
\begin{tabular}{|c|ccccc|}
\hline
$\tau$ & 1 & 5 & 10 & 25 & 50 \\
\hline
$f^{(2)}_{\rm reg}$ & -59.420(6) & -61.242(6) & -61.721(7) & -61.843(8) & -61.553(8)
\\\hline\hline
$f^{(2)}_{\rm lbl}$
& $\D{0.49(1)\atop-0.65(1)\dot\imath}$
& $\D{-0.48(1)\atop-0.72(1)\dot\imath}$
& $\D{-1.10(1)\atop-0.71(1)\dot\imath}$
& $\D{-2.13(1)\atop-0.69(1)\dot\imath}$
& $\D{-3.07(1)\atop-0.68(1)\dot\imath}$
\\\hline\hline
$f^{(2)}_{\rm reg}$ & -59.636(6) & -61.278(6) & -61.716(7) & -61.864(8) & -61.668(8)
\\\hline
$f^{(2)}_{\rm lbl}$
& $\D{0.79(1)\atop-12.45(1)\dot\imath}$
& $\D{-5.61(1)\atop-13.55(1)\dot\imath}$
& $\D{-9.45(1)\atop-13.83(1)\dot\imath}$
& $\D{-15.32(1)\atop-14.03(1)\dot\imath}$
& $\D{-20.26(1)\atop-14.10(1)\dot\imath}$
\\\hline
\end{tabular}
\caption{$f^{(2)}_{\rm reg}(\tau)$ and $f^{(2)}_{\rm lbl}(\tau)$ at some typical values of $\tau$.
The first two rows for $\eta_c$ and the last two for $\eta_b$.
\label{Table:1}}
\end{table}

The $\eta_c$ form factor at zero momentum transfer squared reads
\bqa
\label{F0:concrete:expr:2:loop}
&& F(0) = {e_c^2\over m^{5/2}} \langle \eta_c \vert \psi^\dagger \chi(\mu_\Lambda) \vert 0 \rangle
\Bigg\{ 1 + C_F {\alpha_s(\mu_R)\over \pi}\left( \frac{\pi^2}{8}-\frac{5}{2} \right)
\nn\\
&& + {\alpha_s^2\over \pi^2} \bigg[
 C_F \left({\pi^2\over 8}-{5\over 2} \right) {\beta_0\over 4}\ln {\mu_R^2\over m^2}
- \pi^2 C_F \left(C_F + \frac{C_A}{2} \right) \ln {\mu_{\Lambda}\over m}
\nn\\
&& +f^{(2)}_{\rm reg}(0)+f^{(2)}_{\rm lbl}(0) \bigg]+{\cal O}(\alpha_s^3)\Bigg\},
\eqa
where
\bqa
&& f^{(2)}_{\rm reg}(0) =  -21.10789797(4) C_F^2     -4.79298000(3) C_F C_A \nn\\
&& -\left(\frac{13 \pi ^2}{144}+\frac{2}{3} \ln 2+\frac{7 }{24} \zeta (3) -\frac{41}{36}\right)
  C_F  T_F n_L  \nn\\
 && +  0.223672013(2) C_F T_F  n_H,
\eqa
and
\bqa
&& f^{(2)}_{\rm lbl}(0) =  \bigg(0.73128459+ i\,\pi\left(\frac{\pi ^2}{9}-\frac{5}{3}\right)\bigg) C_F T_F \sum_{i}^{n_L}\frac{e_i^2}{e_Q^2}
\nn\\
 &&+\bigg(0.64696557+2.07357556\,i\bigg)C_F T_F n_H,
\eqa
where $n_L=3$ ($4$) for $e_Q=e_c$ ($e_b$).
As a bonus, Eq.~(\ref{F0:concrete:expr:2:loop}) can be converted into the
order-$\alpha_s^2$ prediction to the $\eta_c \to \gamma \gamma$ partial width, with the aid of the formula
$\Gamma[\eta_c \to \gamma \gamma]={\pi \alpha^2\over 4}|F(0)|^2 M_{\eta_c}^3$.
Our result is consistent with, but much more accurate than,
the previous ${\cal O}(\alpha_s^2)$ result~\cite{Czarnecki:2001zc}.
For completeness, we also include the ``light-by-light" contribution previously
neglected in \cite{Czarnecki:2001zc}.

It is widely known that perturbative expansion for $F(0)$ exhibits poor convergence behavior.
Taking $\mu_\Lambda=m$, the NNLO corrections
overshoot the NLO corrections for both $\eta_c$ and $\eta_b$, but the situation for the $\eta_c$ is provokingly
unsatisfying. The large negative prefactor of $\ln \mu_\Lambda$ in (\ref{F0:concrete:expr:2:loop}) implies
that lowering the NRQCD factorization scale would effectively
dilute the NNLO correction.
In fact, in analyzing ${\cal O}(\alpha_s^2)$ correction to $J/\psi(\Upsilon)\to e^+e^-$~\cite{Beneke:1997jm},
it was suggested that a better choice is $\mu_\Lambda=1$ GeV, the typical bound state scale for both
charmonium and bottomonium. With this scale setting,
$\eta_b\to\gamma\gamma$ receives the accidentally vanishing NNLO correction, but
the convergence of perturbative expansion for $\eta_c$ has not yet been
improved to a satisfactory degree.

\begin{figure}[hbt]
    \centering
    \includegraphics[width=0.4\textwidth]{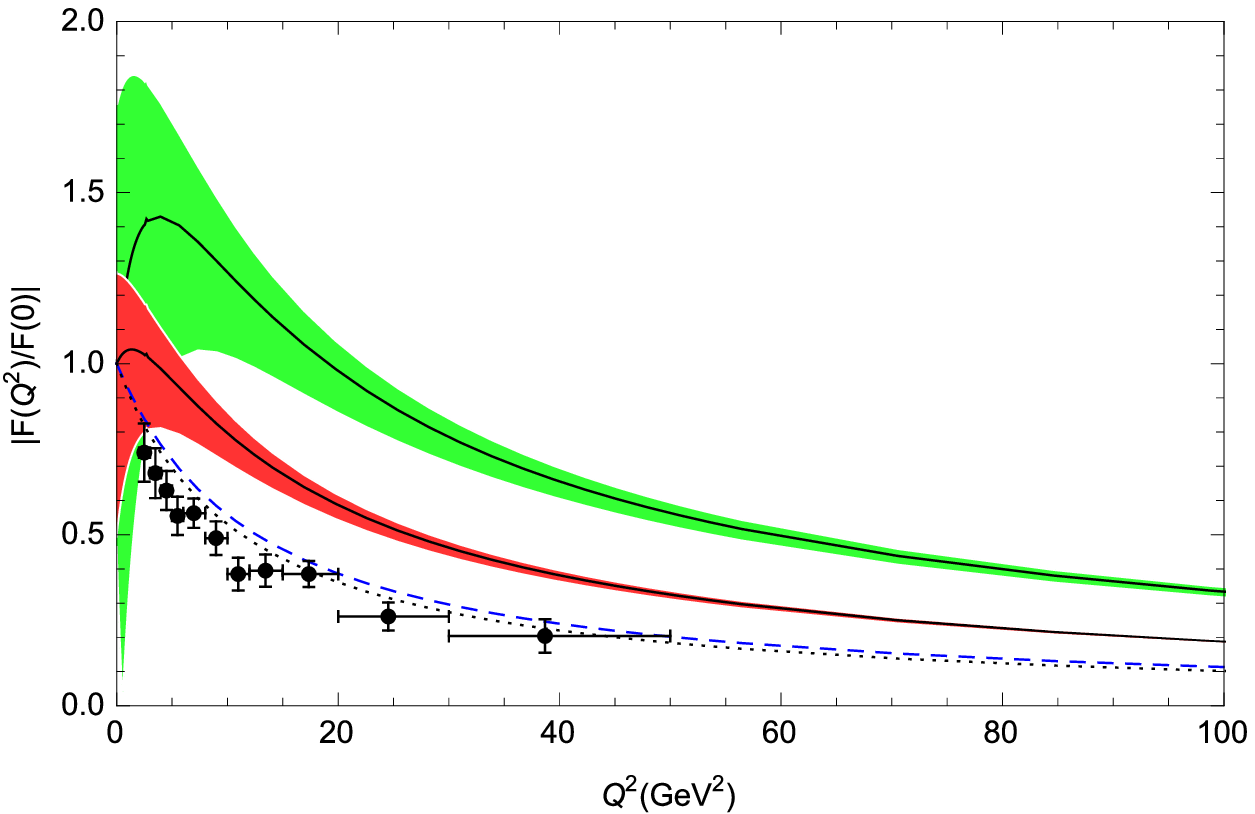}
    \includegraphics[width=0.4\textwidth]{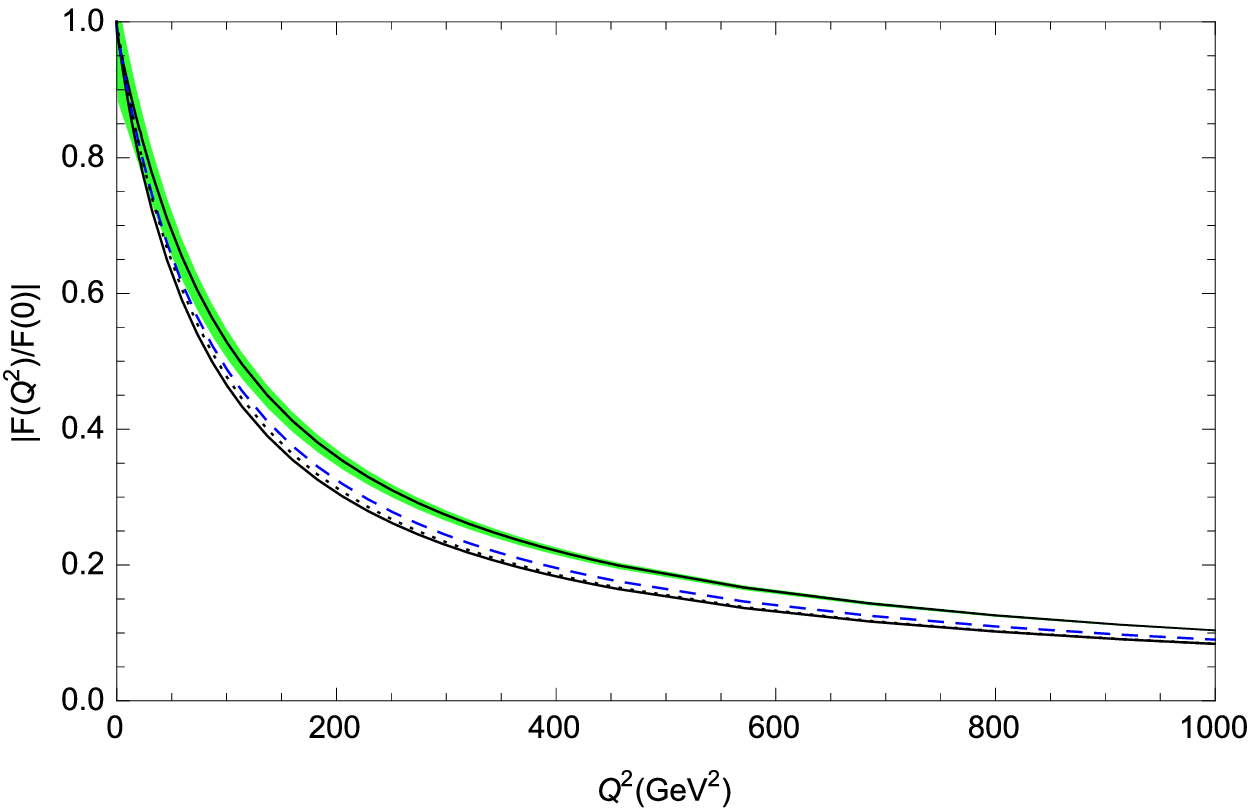}
    \caption{The transition form factor $F(Q^2)$ normalized to $F(0)$ in variation with $Q^2$: $\gamma^*\gamma\eta_c$ (upper); $\gamma^*\gamma\eta_b$ (lower).
    The solid dots with error bars represent the \textsc{BaBar} measurements~\cite{Lees:2010de}.
    The dotted curve stands for the LO NRQCD predictions, the (blue) dashed curve for the NLO predictions,
    while the solid curve for the NNLO predictions.
    The uncertainty bands associated with the NNLO predictions are obtained by varying $\mu_R^2$ from ${1\over 2}(Q^2+m^2)$ to $2(Q^2+m^2)$. There are two distinct NNLO bands, the upper (green) one
    corresponds to $\mu_\Lambda=m$, the lower (red) one to $\mu_\Lambda=1$
    GeV. We suppress the (red) $\mu_R$-uncertainty band for $\eta_b$ since it
    is too narrow to be visible.
    \label{ff}}
\end{figure}

In Fig.~\ref{ff}, we plot the ratio $\vert F(Q^2)/F(0) \vert$ against $Q^2$ over a wide range,
juxtaposing our predictions at various levels of accuracy in $\alpha_s$ with the \textsc{BaBar} data~\cite{Lees:2010de}.
For $F(Q^2)$, we choose $\mu_R=\sqrt{Q^2+m^2}$ in (\ref{NRQCD:short-dist:coef:2:loop}) by default;
for $F(0)$ in (\ref{F0:concrete:expr:2:loop}), we fix $\mu_R=m$.
Using the latest PDG determination of the $\overline{\rm MS}$ masses $\overline{m}_c(\overline{m}_c)=1.28$ GeV and $\overline{m}_b(\overline{m}_b)=4.18$ GeV~\cite{Agashe:2014kda}, we convert them into
the 2-loop quark pole masses as $m_c=1.68$ GeV and $m_b=4.78$ GeV with the aid of
the package \textsf{RunDec}~\cite{Chetyrkin:2000yt}.
We also use \textsf{RunDec} to evaluate the running $\alpha_s^{\overline{\rm MS}}$.
To assess the uncertainty induced by the renormalization scale,
we slide $\mu_R$ from $1/\sqrt{2}$ to $\sqrt{2}$ times the default choice.
For completeness, we also include the ``light-by-light" scattering contribution in our phenomenological analysis.

As expected, the LO NRQCD prediction describes the measured $\gamma^*\gamma\eta_c$ form factor quite well.
The agreement survives after adding the modest negative NLO perturbative correction.
Since the NLO prediction is stable against the variation of $\mu_R$, we suppress its uncertainty band
in Fig.~\ref{ff}.

It is counterintuitive that the NNLO prediction
appears to have stronger $\mu_R$-dependence with respect to the NLO result. This symptom can be attributed to the unusually large $|f^{(2)}(\tau)|$.
The NNLO prediction turns out to be also sensitive to the
NRQCD factorization scale $\mu_\Lambda$ (This is mainly because we keep $F(Q^2)/F(0)$ unexpanded).
To discern the uncertainty induced by $\mu_\Lambda$,
we include two sets of NNLO predictions in Fig.~\ref{ff},
one with $\mu_\Lambda=m$ (the ultimate UV cutoff scale of NRQCD) and
$\mu_\Lambda=1$ GeV (roughly the scale of $m v$ for charmonium and bottomonium).
From the discussions following (\ref{F0:concrete:expr:2:loop}),
we see that the former choice results in rather poor convergence for perturbative series.
Indeed, it yields a prediction to $\vert F(Q^2)/F(0) \vert$
that severely deviates from the \textsc{BaBar} data over the experimentally
accessible $Q^2$ range.
For the latter scale setting, the discrepancy is considerably
relieved, yet still pronounced, $5\sigma$ away from most data points.
To reconcile the NRQCD prediction with the data, it is tempting to further decrease the value of $\mu_\Lambda$,
however one then faces the dilemma that NRQCD factorization would
cease to make sense.

In passing, it is worth mentioning that there exist different ways to ``interpret" the
NNLO corrections to $F(Q^2)/F(0)$, {\it i.e.},
 to expand this ratio strictly as power series in $\alpha_s$ through the 2nd order,
 or take the renormalization scales associated with $F(Q^2)$ and $F(0)$ interchangeably.
In some scenarios, the NNLO correction turns out to be tiny and there reaches a perfect agreement between
the ``predictions" and data.
A comprehensive survey reveals that, keeping $F(Q^2)/F(0)$ expanded or not is not the key that influences the
agreement, rather, the key is rooted in whether one takes two $\mu_R$s entering $F(Q^2)$ and $F(0)$ equal or not.
As a consequence of the deeply negative and flat $f^{(2)}(\tau)$, we find those predictions with equal-$\mu_R$ recipe
will always yield decent agreement with the data (no matter expanding $F(Q^2)/F(0)$ or not),
those with unequal $\mu_R$ (like in our case) will generally result in severe discrepancy between the theory and the data.
Note that, unlike the ratio of the $S$-wave quarkonium decay rates as studied in \cite{Czarnecki:2001zc,Kiyo:2010jm},
it is quite unnatural to take the $\mu_R$ in $\eta_c$ production process equal to that in the $\eta_c$ decay.
They rather should be taken as the characteristic hard scale in respective production/decay processes.
Moreover, since the experimental data are available separately for $F(Q^2)$ and $F(0)$, we
 can always convert the measured $\eta_c\to \gamma\gamma$ width to $|F(0)|$, then extract the NRQCD matrix element to the NNLO accuracy, then plug it to (\ref{NRQCD:factorization}) to predict the $F(Q^2)$ over a wide range of $Q^2$.
 In computing the ratio $F(Q^2)/F(0)$, we would reproduce exactly the same plot as in Fig.~\ref{ff}.
 In our opinion, Fig.~\ref{ff} remains the most honest way to present our NNLO prediction.
It is important to note that the {\it rise-then-drop} shape of our prediction to $F(Q^2)$ is qualitatively
incompatible with the monopole shape observed experimentally.

For $\gamma^*\gamma\to \eta_b$, the size of NNLO correction to $|F(Q^2)/F(0)|$ seems to be modest even at
$\mu_\Lambda= m_b$, and accidentally small for $\mu_\Lambda= 1$ GeV. Thus, the NRQCD factorization appears to work
reasonably well for bottomonium. One has to await the next generation of high-energy $e^+e^-$
collider to test our predictions.

We tend to conclude that, within the reasonable ranges of $\mu_\Lambda$ and $\mu_R$,
the state-of-the-art NRQCD prediction utterly fails to account for the \textsc{BaBar} data on
the $\gamma^*\gamma\to \eta_c$ form factor. Since our predictions do not involve any adjustable
nonperturbative parameters, and the disagreement prevails for a wide range of data points,
this discrepancy poses a much more thorny challenge to theory than what was encountered
in the $e^+e^-\to J/\psi+\eta_c$~\cite{Braaten:2002fi,Liu:2002wq,Hagiwara:2003cw}.

A pressing question is how to resolve this disturbing puzzle within the confine of NRQCD.
The relativistic correction to this transition form factor appears to be modest.
Certainly it would be extremely illuminating to know the actual size of ${\cal O}(\alpha_s^3)$ correction,
provided such a calculation were feasible in the foreseeable future.
Furthermore, as was noted in  \cite{Jia:2008ep,Jia:2010fw},
an outstanding shortcoming of fixed-order NRQCD calculation for quarkonium production
is that, since the NRQCD matching coefficient contains several distinct scales $Q$, $m$, $\mu_\Lambda$,
it is {\it a priori} unclear how to optimally assign these scales to various $\alpha_s$.
Without getting rid of this inherent scale ambiguity,
it is difficult to make a sharp higher-order prediction out of the
NRQCD framework.

If the future \textsc{Belle} measurement of the $\gamma\gamma^*\to \eta_c$ transition form factor
confirms the \textsc{BaBar} data, and if this discrepancy persists, serious doubt will be cast onto
the usefulness of the NRQCD factorization approach when applied to
charmonium exclusive production processes.

\begin{acknowledgments}
{\noindent\it Acknowledgment.} We thank Shuang-Ran Liang for assistance in obtaining Eqs.~(8) and (9).
The work of F.~F. is supported by the National Natural Science Foundation of China under Grant No.~11505285,
and by the Fundamental Research Funds for the Central Universities.
The work of Y.~J. is supported in part by the National Natural Science Foundation of China under Grants No.~11475188,
No.~11261130311 (CRC110 by DGF and NSFC), by the IHEP Innovation Grant under contract number Y4545170Y2,
and by the State Key Lab for Electronics and Particle Detectors.
W.-L.~S. is supported by the National Natural Science Foundation of China under Grant No.~11447031, and
by the Open  Project Program of State Key Laboratory of Theoretical Physics under Grant No.~Y4KF081CJ1,
and also by the Fundamental Research Funds for the Central Universities under Grant No. SWU114003, No. XDJK2016C067.
\end{acknowledgments}

\end{document}